\documentstyle[epsfig,aps,prl,twocolumn,doublespace,12pt]{revtex}
\begin{document}
%
%
%
%
\onecolumn
\title{The Pauli Exclusion Principle and $SU(2)$ vs. $SO(3)$ in Loop Quantum
Gravity}
\author{John Swain}
\address{Department of Physics, Northeastern University, Boston, MA 02115, USA\\
email: john.swain@cern.ch}
\date{Submitted for the Gravity Research Foundation Essay Competition, March 27,
2003}
\maketitle

\begin{abstract}
\section*{\bf Abstract}

Recent attempts to resolve the ambiguity in the loop quantum
gravity description of the quantization of area has led to 
the idea that $j=1$ edges of spin-networks dominate in 
their contribution to black hole areas as opposed to $j=1/2$ which
would naively be expected. This suggests that
the true gauge group involved might be $SO(3)$ rather than $SU(2)$
with attendant difficulties.
We argue that the assumption that a
version of the Pauli principle is present in loop quantum gravity
allows one to maintain $SU(2)$ as the gauge group while still 
naturally achieving the desired suppression of spin-1/2 punctures.
Areas come from $j=1$ punctures rather than $j=1/2$ punctures
for much the same reason that photons lead to macroscopic classically
observable fields while electrons do not.
\end{abstract}
%
%
\section{Introduction}
The recent successes of the approach to canonical quantum gravity
using the Ashtekar variables have been numerous and significant. Among
them are the proofs that area and volume operators have discrete
spectra, and a derivation of black hole entropy up to an overall
undetermined constant\cite{Ashtekarstuff}. An excellent recent review
leading directly to this paper is by Baez\cite{Baez}, and
its influence on this introduction will be clear.

The basic idea is that a basis for the solution of
the quantum constraint equations is given by {\em spin-network} states,
which are graphs whose edges carry representations $j$ of $SU(2)$.
To a good approximation, the area $A$ of a surface which intersects
a spin network at $i$ edges, each carrying an $SU(2)$ label $j$ is given
in geometrized units (Planck length equal to unity) by
\begin{equation}
A \approx \sum_i 8\pi\gamma\sqrt{j_i(j_i+1)}
\end{equation}
where $\gamma$ is the Immirzi-Barbero parameter\cite{Immirzi-Barbero}.
The most important microstates 
consistent with a given area are those for which $j$ is as small as possible,
which one would expect to be $j_{min}=1/2$. In this case, each contribution
to the area corresponds to a spin $j=1/2$ which can come in two possible
$m$ values of $\pm 1/2$. For $n$ punctures, we have
$A \approx 4\pi\sqrt{3}\gamma n$ and entropy $S\approx \ln(2^n) \approx
\frac{\ln(2)}{4\pi\sqrt{3}\gamma}A$. 

Now looking outside loop quantum gravity for help, we can use
Hawking's formula\cite{Hawking} for black hole entropy $S=A/4$ to get
$\gamma=\frac{\ln(2)}{\pi\sqrt{3}}$ and the smallest quantum of 
area is then $8\pi\gamma\sqrt{\frac{1}{2}(\frac{1}{2}+1)} = 4\ln(2)$.
Physically this is very nice as it says that a black hole's horizon
acquires area, to a good approximation, from the punctures of many
spin network edges, each carrying a quantum of area $4\ln(2)$ and
one ``bit'' of information -- a vindication Wheeler's ``it from bit''
philosophy\cite{Wheeler}.

Bekenstein's early intuition\cite{Beckenstein} that the area operator
for black holes should have a discrete spectrum made of equal area steps
(something not really quite true in loop quantum gravity in full generality)
was followed by Mukhanov's observation \cite{Mukhanov} that 
the $n^{th}$ area state should have degeneracy $k^n$ with
steps between areas of $4\ln(k)$ for $k$ some integer $\geq 2$ 
in order to reproduce the Hawking expression $S=A/4$. For $k=2$ one would have
the $n^{th}$ area state described by $n$ binary {\em bits}. 

On the other hand,  Hod\cite{Hod} has argued that by looking at the
quasinormal damped modes of a classical back hole one should be
able to derive the quanta of area in a rather different way. The basic
idea is to use the formula $A=16\pi M^2$ relating area and mass of a
black hole to get $\Delta A=32\pi M \Delta M$ for the change in area 
accompanying an emission of energy $\Delta M$. Nollert's computer
calculations\cite{Nollert}
of the asymptotic frequency $\omega$ of the damped normal modes gave
$\omega \approx 0.4371235/M$, so setting $\omega = \Delta M$ one finds
$\Delta A \approx 4.39444$. It is tempting then to conclude that perhaps
$\Delta A = 4\ln(3)$. Motl \cite{Motl} later showed
that this is indeed correct, and not just a fortuitous numerical coincidence.

Since then, Dreyer\cite{Dreyer} has pointed out that one might well
expect $\Delta A \approx 4\ln(3)$ instead of $\Delta A \approx 4\ln(2)$
if the spin network edges contributing to the area of a black
hole didn't carry $j=1/2$, but rather $j=1$. In this case
$j_{min}$ would be $1$ rather than $1/2$,
there would be
three possible $m$ values, and area elements would be described not by
binary ``bits'', but by trinary ``trits''. (See also \cite{MotlandNeitzke}).
This also suggests
that perhaps the correct gauge group is not $SU(2)$ but $SO(3)$, although
this could complicate the inclusion of fermions in the theory.

Corichi has recently argued\cite{Corichi} that one might arrive at
the conclusion that $j_{min}=1$ by suggesting that one should think of
a conserved fermion number being assigned to each spin-1/2 edge.
Adding or losing an edge's worth of area would have to mean
that at some point a spin-1/2 edge would be essentially dangling
in the bulk ({\it i.e.} not imbuing the horizon surface with area)
and this should not be allowed. If edges carried $j=1$ 
one could imagine coupling the edge to a fermion-antifermion pair
and this would locally solve the fermion number problem. This is 
quite appealing as one might then think of the loss of an element of
area with accompanying fermion-antifermion production in Hawking
radiation as the detachment of a spin-1 edge from the horizon
which then couples to an $f\bar{f}$ pair. As Corichi\cite{Corichi} points out:

{\em ``the existence of $j=1/2$ edges puncturing the horizon is not forbidden
\ldots, but they must be suppressed. Thus, one needs a dynamical 
explanation of how exactly the entropy contribution is dominated by
the edges with the dynamical allowed value, namely $j=1$.''}

\section{The Exclusion Principle}

The point of this essay is to suggest that one 
might want to 
assume that
a version of the spin-statistics theorem (or, equivalently, the
Pauli exclusion principle)
applies to loop quantum gravity. More precisely, it could be the
case that no more than two punctures of $j=1/2$, each with differing
$m$ values, may puncture a given surface. In this case,
the dominance of $j=1$ punctures
(even though $j=1/2$ is allowed) is very natural: if only a maximum
of two spin-1/2 edges can puncture any surface then for large
numbers of punctures one would have
an effective $j_{min}=1$ despite the gauge group being $SU(2)$.

The spin-statistics theorem as usually formulated and proven
(to the extent that one rigorously proves anything in quantum field
theory!), is, of course,
for matter fields in a background spacetime usually assumed to be
flat. It is not entirely clear what sort of extension should apply to 
amplitudes in quantum gravity. 

On the other hand, the spin-statistics association is strongly
combinatorial in flavour and seems natural in a spin-network context.
Surely for a surface punctured by $n$ edges it would be natural to 
associate an amplitude which returns to its original value, up to 
a phase, upon the exchange of two spin-1/2 (and thus identical,
indistinguishable) punctures. If making the exchange twice leads to 
the identity\footnote{It is interesting to consider the possibility of
more exotic braid or anyon-like statistics if one would have
to keep track of how one edge moved around another, but this is
beyond the scope of this essay.}, one then needs merely to
choose a sign, and -1 seems at least as natural as +1.  In the usual
quantum field theory one simply tries each possibility, finding that
things only work out when the -1 is applied to half integral spin
fields and the +1 to integral spin fields.

In loop quantum gravity this leads
then to a picture in which a black hole {\em can} get area
contributions from
spin-1/2 (and spin-3/2, spin-5/2, {\em etc.}) punctures, but these
are always very small compared to the enormous number of $j=1$
edges. The value $j=1$ is the lowest value of $j$ contributing
nonzero area not
being severely limited by Fermi-Dirac statistics, and able to 
appear arbitrarily often.

In a sense, the question of $SU(2)$ vs. $SO(3)$ in 
loop quantum gravity could be
very much like one that we face in everyday physics. Integer spin particles,
which fall into $SO(3)$ representations, obey Bose-Einstein statistics and
gregariously bunch together to give large macroscopically observable
fields such as electromagnetic fields.  Half-integer spin particles do not.
We could well be excused
for thinking 
that the symmetry group of our world under rotations was $SO(3)$
rather than $SU(2)$. Indeed, until the discovery of spin, it did appear
that physical rotations were always elements of $SO(3)$. The
need for $SU(2)$ was, in many ways, a surprise!

It may be hard to find direct experimental evidence of these ideas,
but it is at least possible to make some predictions.
For example, the $SU(2)$ theory with the exclusion principle proposed
here will give both 

\begin{itemize}
\item[a)] what seems to be the correct result for large
black holes, with areas well-described by values which go up {\em in
steps of} $4\ln(3)$; {\underline{and}}
\item[b)] the possibility
simultaneously admitting areas {\em as small as} $4\ln(2)$.
\end{itemize}

\section*{Acknowledgements}

I would like to thank the NSF for their continued 
and generous support.

\end{document}